# OMSN and FAROS: OCTA Microstructure Segmentation Network and Fully Annotated Retinal OCTA Segmentation Dataset

Peng Xiao, Xiaodong Hu, Ke Ma, Gengyuan Wang, Ziqing Feng, Yuancong Huang, and Jin Yuan

***Abstract*—The lack of efficient segmentation methods and fully-labeled datasets limits the comprehensive assessment of optical coherence tomography angiography (OCTA) microstructures like retinal vessel network (RVN) and foveal avascular zone (FAZ), which are of great value in ophthalmic and systematic diseases evaluation. Here, we introduce an innovative OCTA microstructure segmentation network (OMSN) by combining an encoder-decoder-based architecture with multi-scale skip connections and the split-attention-based residual network ResNeSt, paying specific attention to OCTA microstructural features while facilitating better model convergence and feature representations. The proposed OMSN achieves excellent single/multi-task performances for RVN or/and FAZ segmentation. Especially, the evaluation metrics on multi-task models outperform single-task models on the same dataset. On this basis, a fully annotated retinal OCTA segmentation (FAROS) dataset is constructed semi-automatically, filling the vacancy of a pixel-level fully-labeled OCTA dataset. OMSN multi-task segmentation model retrained with FAROS further certifies its outstanding accuracy for simultaneous RVN and FAZ segmentation.***

## I. INTRODUCTION

THE evaluation of retinal vasculature not only is essential for understanding the disease progression of many kinds of ophthalmopathy and retina-related systemic diseases [1]–[5] but also plays a significant role in clinical diagnosis and treatment [5]–[9]. Optical coherence tomography (OCT), a famous optical imaging technique for observing the retinal vasculature, is non-invasive and can present comprehensive information on retinal structures, but it cannot provide the functional information of blood flow [10]. In contrast, OCT angiography (OCTA), an OCT-based functional imaging technique, uses coherent light to photograph the fundus, thus accomplishing a non-invasive, label-free solution for imaging the retinal vasculature with capillary-level resolution [11]. Hence, OCTA not only provides a better chance of revealing the subtle microvascular distortions associated with early asymptomatic ocular disease, such as blood vessel excretion, abnormal fovea, increased blood vessel tortuosity [12], but also possesses the ability to show both structural and blood flow information in tandem, with which ophthalmologists are facilitated to localize and delineate pathology specifically [13].

Several studies have demonstrated that the segmentation and quantitative assessment of OCTA microstructures like retinal vessel network (RVN) and foveal avascular zone (FAZ) are of great significance for disease diagnosis and treatment [3], [5], [9]. However, massive screening for common ocular conditions calls for efficient automatic segmentation of OCTA images, since manual delineation is extremely complex and heavily reliant on sophisticated ophthalmologists. Recently, many conventional machine learning (ML) methods and deep learning (DL) methods have been developed to segment RVN or FAZ in OCTA images respectively [3], [10], [14]–[23]. Besides, several publicly available OCTA segmentation datasets with partial RVN and FAZ annotations are released to researchers for further study of OCTA images segmentation [3], [24], [25]. To our best knowledge, although many advances have been made in the research on automatic segmentation of OCTA images, it still faces two great challenges: the lack of advanced algorithms to automatically segment RVN and FAZ simultaneously, and the scarcity of fully annotated OCTA datasets.

Therefore, we first innovatively propose an OCTA Microstructure Segmentation Network (OMSN) to achieve concurrent and effective RVN and FAZ segmentation. The proposed OMSN (Fig. 1a) mainly consists of a multi-scale skip connection encoder-decoder architecture, the novel split-attention on-based ResNeSt blocks [26], and a few convolutional units. The fundamental framework of OMSN is the encoder-decoder structure with multi-scale skip connections, which is inspired by Unet 3+ [27]. The core of OMSN is the novel split-attention-based backbone network, ResNeSt [26], which is a novel variant of ResNet [28]. Secondly, we build and release a Fully Annotated Retinal OCTA Segmentation (FAROS) dataset (Fig. 3c) to address the problem of the

†This work was supported in part by the National Natural Science Foundation of China (Nos. 82230033 and 82271133), Guangzhou Science and Technology Program (No. 202103000043), Basic and Applied Basic Research Foundation of Guangdong Province (No. 2022A1515011486), and Guangzhou Science and Technology Plan Project (No. 202103010001). (P. Xiao, X. Hu and K. Ma contributed equally to this work.) (Corresponding authors: J. Yuan, yuanjincornea@126.com; P. Xiao, xiaopengaddis@hotmail.com; X. Hu, huxiaodong@gzzoc.com)

P. Xiao, X. Hu, K. Ma, G. Wang, Z. Feng, Y. Huang, and J. Yuan are with State Key Laboratory of Ophthalmology, Zhongshan Ophthalmic Center, Sun Yat-sen University, Guangdong Provincial Key Laboratory of Ophthalmology and Visual Science, Guangzhou 510060, China.

scarcity of fully annotated OCTA datasets. Considering the fact that annotating RVN precisely demands massive human and time resources, we adopt a semi-automatic scheme to obtain pixel-level RVN including capillaries (RVC) and FAZ annotations.

With thorough experiments, the proposed OMSN is demonstrated to achieve outstanding performance in two single segmentation tasks (RVN, FAZ) over two open datasets, ROSE-1 and OCTA_3M, respectively. Besides, we achieve excellent simultaneous multi-task OCTA segmentation (RVN and FAZ) based on a single OCTA image and a multi-class model on the two publicly available datasets and the FAROS dataset for the first time. Especially, we quantitatively demonstrate the concurrent RVN and FAZ segmentation models outperform the single-task segmentation models built on the same datasets.

To summarize, our contributions mainly include the following parts:

1) We introduce an innovative segmentation network, OMSN, for extracting microstructures (RVN or/and FAZ) in OCTA images. Comparative experiments prove that our method achieves excellent RVN or/and FAZ segmentation in OCTA images.

2) We groundbreakingly build and publish a fully-labeled OCTA dataset, FAROS [1], which contains pixel-level RVC and FAZ annotations annotated by a semi-automatic method.

3) We retrain a multi-task segmentation model using OMSN on the FAROS dataset, and the internal testing results certify that it reaches high segmentation performance.

## II. RELATED WORKS

### A. ML-based segmentation methods

ML-based methods generally utilize digital image processing and feature extraction to obtain the segmentation of OCTA images [29]. Several studies have explored ML-based RVN and FAZ segmentation method to quantify and standardize retinal vascular changes, respectively. For example, Gao *et al*. [14] applied split-spectrum amplitude-decorrelation angiography (SSADA) algorithm to OCTA images, then binarized them according to a decorrelation threshold. Eladawi *et al*. [15] employed a joint Markov-Gibbs random field model to segment RVN in OCTA images. Li *et al*. [16] proposed an algorithm based on anisotropic diffusion filter, top-hat filter and global threshold to get RVN segmentation in OCTA images. Lu *et al*. [17] combined region growing algorithm and morphologic operations, supplemented by an innovative generalized gradient vector flow (GGVF) model to realize FAZ segmentation in OCTA images. Diaz *et al*. [18] introduced white top hat operation to improve image contrast, thus obtaining good performance of FAZ segmentation by means of vascularity edge identification.

### B. DL-based segmentation methods

Derived from ML, DL-based methods with large amounts of [30]. To realize precise pixel-level semantic segmentation, fully convolutional networks (FCN) [31] replaces the full connection layer with corresponding convolutional layer in convolutional neural networks (CNN), such that a typical encoder-decoder structure is formed. While FCN adopts a point-by-point summation method in the skip connection layer, Unet [32] utilizes feature channel dimension concatenation. Since its proposal, Unet has become the best-known network in the field of medical image segmentation for its concise and practical structure and support for a small amount of data. For example, Pissas *et al*. [23] combined two modified Unet to form a iterative refinement segmentation network, iUnet, and realized good RV segmentation results in OCTA images. Guo *et al*. [19] utilized a deeper U-shaped network to obtain a considerably good FAZ segmentation in OCTA images.

In addition to representative structures, several creative networks with novel structures or modules have enriched the field of RVN and FAZ segmentation. For instance, Ma *et al*. [3] introduced an OCTA vessel segmentation model with split-based coarse-to-fine structure. The U-shaped coarse stage was used to obtain a preliminary confidence map of vessels, and then a fine stage was applied to tune the contour of the retinal vasculature. Li *et al*. [10] put forward a 3D-to-2D projection network for 3D OCT and OCTA data whose core was a projection learning module. Wu *et al*. [33] proposed a novel SCS-Net to accurately segment RVN within variant scales and complicated anatomies. Mou *et al*. [34] proposed an innovative channel and spatial attention network (CS-Net) that was suitable for several curvilinear structure segmentation. Mou *et al*. [21] upgraded CS-Net and extended the application scenarios from 2D to 3D which covered many curvilinear organ structures. To address the limited samples and contrast artifacts, Liu *et al*. [35] proposed the artifacts and contrast robust representation of OCTA semi-supervised segmentation (ACRROSS), using a conditioning network (Unet or CS-Net) to learn anatomy composition to accurately segment vessels and capillaries. Hao *et al*. [36] proposed a voting-based adaptive feature fusion (VAFF) multi-task network for the joint segmentation, detection, and classification of RVN, FAZ, and retinal vascular junction (RVJ) in three different OCTA *en face* angiograms. Differently, we propose a multi-class segmentation model based on a single OCTA image to simplify the input and improve the multi-task segmentation performance.

## III. METHODS

The methods applied in this study mainly include the architecture of OMSN, processing for OMSN input and output, and loss functions for single and multi-task segmentation (Fig. 1a).

### A. Network Architecture

The pipeline of our proposed OMSN principally consists of a multi-scale skip connection encoder-decoder framework, ResNeSt blocks, and a few convolutional units.

---

[1] https://www.zenodo.org/record/6381668#.Y5LqN8tByUk

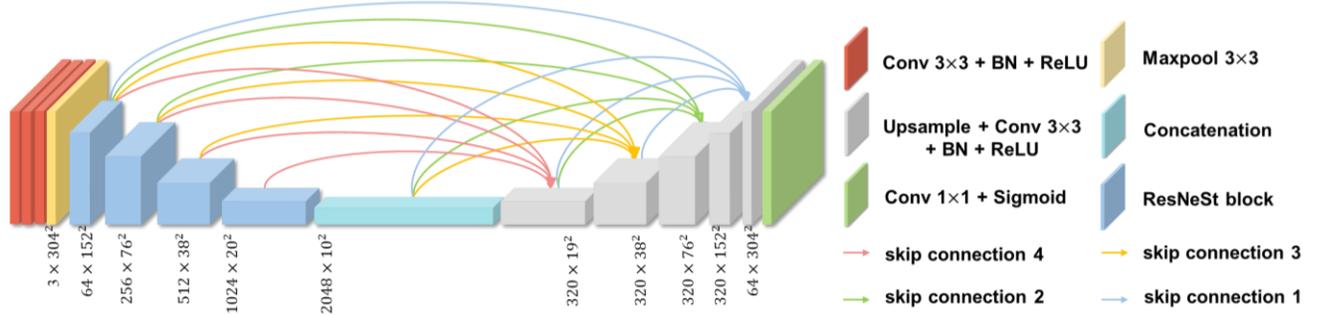

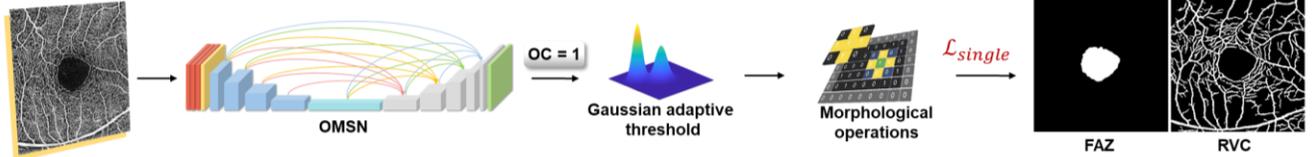

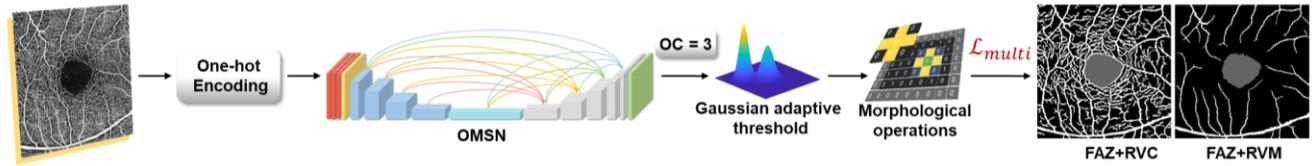

Fig. 1. The proposed OMSN and the experimental flows in this paper. **a** The architecture of our proposed OMSN, which principally consists of input convolutional layers, multi-scale skip connection encoder-decoder architecture, and ResNeSt blocks. **b** The flow of single-task segmentation. **c** The flow of multi-task segmentation. OC: output channel.

*1) The multi-scale skip connection encoder-decoder:* As illustrated in Fig. 1a, when an input image $X_{in}$ comes into the OMSN network, there follows a fundamental feature extraction unit consisting of a convolutional layer, batch normalization (BN), activation function ReLU, and a max pooling layer. The first feature map $X_{en}^1$ is computed as:

$$X_{en}^1 = maxpooling(\mathcal{G}(X_{in})) \quad (1)$$

where $\mathcal{G}(\cdot)$ is convolution operation followed with BN and ReLU, and maxpool($\cdot$) represents max pooling operation.

To explicitly merge superficial and deep semantic information, the following part aggregates multi-scale features with multi-scale skip connections. The multi-scale skip connections combine a typical encoder-decoder architecture with multi-scale feature concatenation. The encoder module achieves downsampling operation by ResNeSt encoder blocks. The decoder module realizes upsampling operation by utilizing bilinear interpolation, convolution, BN, and ReLU function. Multi-scale skip connections concatenate multi-scale feature maps by down skip connections, parallel skip connections, and up skip connections respectively. Down skip connections deliver the high-level semantic information from the shallower layers by ResNeSt encoder blocks. Parallel skip connections link the feature maps within the same level. Up skip connections transmit the low-level information from the deeper layers by upsampling operation. It is noted that appropriate zero-padding and truncation operations are used in the process of downsampling and upsampling because the input image resolution is $304 \times 304$. As shown by skip connection 4 in Fig. 1a, five feature maps with a size of (64,19,19) are generated by three down skip connections, a parallel skip connection, and an up skip connection, and then concatenated to form a feature map with a size of (320,19,19). Subsequently, a convolution operation followed with BN and a ReLU function is employed to aggregate the multi-channel feature information. Similar processes occur in the other skip connection processes, such as skip connections 1, 2, and 3, and the multi-scale skip connection process can be expressed mathematically as:

$$\begin{cases} X_{de}^i = \mathcal{G}([\underbrace{\mathcal{C}(\mathcal{D}(X_{en}^k))_{k=1}^{i-1}}_{1^{th} \sim i^{th} \, stages}, \mathcal{C}(X_{en}^i), \underbrace{\mathcal{C}(\mathcal{U}(X_{de}^k))_{k=i+1}^{N}}_{1^{th} \sim i^{th} \, stages}]) \\ i = 1, \dots, N-1 \end{cases} \quad (2)$$

where $i$ indexes the decoder layer number, $N$ refers to the total number of network layers, $\mathcal{C}(\cdot)$ evinces convolution operation, [·] refers to concatenation operation, and $\mathcal{D}(\cdot)$ and $\mathcal{U}(\cdot)$ denotes downsampling and upsampling respectively.

After that, feature map $X_{de}^1$ experiences upsampling, convolution, BN, and ReLU function successively, and reverts to the original image resolution with multiple channels. Ultimately, a convolution operation (with OC = 1 or 3 in single-task or multi-task segmentation) and a sigmoid function $Sigmoid(\cdot)$ are utilized to obtain the segmentation results $X_{out}$. This procedure can be computed as:

$$X_{out} = Sigmoid\left(\mathcal{C}(\mathcal{G}(\mathcal{U}(X_{de}^1)))\right) \quad (3)$$

After the sigmoid function, we get a probability distribution map $X_{out}$ with a size of (304,304,1) for single tasks and (304,304,3) for multiple tasks. Each probability value in $X_{out}$ corresponds to the likelihood that the pix belongs to the target

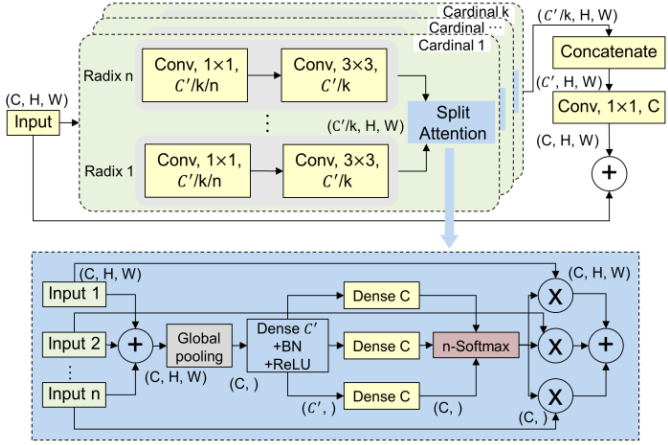

Fig. 2. ResNeSt block with spit-attention module.

category or background.

*2) ResNeSt block with split-attention module:* As shown in Fig. 2, the input of ResNeSt block is a typical residual network block with split-attention modules. The input X of the block is fed into k cardinal groups with exactly the same architecture. In each cardinal, the cardinal group is further equivalently split into n parallel radixes. Every radix is composed of $1 \times 1$ and $3 \times 3$ convolutional layers, and outputs a feature map with the size of $(C'/k, H, W)$. Afterward, a spit-attention module is employed to aggregate the feature maps from n radixes, the details of which are illustrated in the light blue area. This module firstly applies an element-wise summation to fuse n feature maps, and then adopts global pooling to get radix-wise statistics s:

$$s = \frac{1}{H \times W} \sum_{i=1}^{H} \sum_{j=1}^{W} R_a(i,j), a = 1, \ldots, n \quad (4)$$

where $R(i,j)$ refers to the output of every radix.

The subsequent dense blocks and n-Softmax layer are used to obtain weight vectors ω based on s. Then the output U of the spit-attention module is a summation of the weighted input feature maps, which is the consequence of the radix-wise product between ω and R. Finally, the outputs of $k$ cardinal are concatenated together and then go through convolution with kernel size = 1. Formulaically, the out of the ResNeSt blocks Y can be expressed as:

$$Y = \mathcal{F}([U_1, \ldots, U_k]) + \mathcal{T}(X) \quad (5)$$

where $\mathcal{F}(\cdot)$ is a $1 \times 1$ convolution, and $\mathcal{T}(\cdot)$ stands for appropriate transformation, such as combined convolution with pooling or stride convolution.

### B. Processing for OMSN input and output

*1) Gaussian adaptive threshold:* We first turn $X_{out}$ into grayscale images. Then we calculate the local thresholds by the weighted sum of neighborhood values with a specific size where weights are a gaussian window, according to the gray value distribution of different areas of the image.

*2) Morphological operations:* We traverse the entire image with the 8-adjacency principle to find discrete small objects and contiguous small holes, which are smaller than a specified size. After that, we remove the small objects with the method of filling with background value and filter the small holes by means of filling with a target value.

*3) One-hot encoding:* One-hot encoding uses an N-bit state register to encode N states, each state has its own independent register bit, and only one bit is valid at any time. This first requires mapping categorical values to integer values. Then, each integer value is represented as a binary vector.

### C. Loss functions

*1) Single-task loss function:* We empirically combine focal loss [37] $\mathcal{L}_{focal}$ and mean square error (MSE) loss $\mathcal{L}_{MSE}$ to form the single-task loss function. The purpose of focal loss is to give more weight to hard-to-classify samples, which can be formulated as:

$$\mathcal{L}_{focal} = -\alpha_t(1 - p_t)^\gamma \log(p_t) \quad (6)$$

where $p_t$ is a classification probability, $\alpha_t$ is a balanced variant, and γ is a tunable focusing parameter. Based on multiple pre-experiments, $\alpha_t$ and γ are set to 0.6 and 1.2, respectively.

$\mathcal{L}_{MSE}$ is the sum of squared distances between our target variable and predicted values. Then the employed single-task loss function $\mathcal{L}_{single}$ is expressed below:

$$\mathcal{L}_{single} = \mathcal{L}_{focal} + \mathcal{L}_{MSE} \quad (7)$$

*2) Multi-task loss function:* To enhance the subtle RVN and FAZ segmentation detail, we experientially design a multi-task loss function composed of Lovasz-Softmax (LS) loss [38] $\mathcal{L}_{LS}$ and Cross Entropy (CE) loss $\mathcal{L}_{CE}$. $\mathcal{L}_{LS}$ is a loss function for multiclass semantic segmentation that incorporates the softmax operation in the Lovasz extension. The Lovasz extension is a means by which we can achieve direct optimization of the mean intersection-over-union loss in neural networks. $\mathcal{L}_{CE}$ measures the performance of a classification model whose output is a probability value between 0 and 1, and increases as the predicted probability dive from the actual label, which is formulated as:

$$\begin{cases} \mathcal{L}_{CE} = -\frac{1}{N}\sum_j \sum_{c=1}^{M} y_{jc} \log(p_{jc}) \\ y_{jc} = \begin{cases} 1, & j = c \\ 0, & j \neq c \end{cases} \end{cases} \quad (8)$$

where $M$ is the number of categories, j is a sample observation value, c is a category value, $p_{jc}$ is the probability that j belongs to c, and $y_{jc}$ is a symbolic function.

By combining LS loss $\mathcal{L}_{LS}$ and CE loss $\mathcal{L}_{CE}$, the mixed multi-task loss function $\mathcal{L}_{multi}$ is defined as:

$$\mathcal{L}_{multi} = \alpha \mathcal{L}_{LS} + (1 - \alpha)\mathcal{L}_{CE} \quad (9)$$

where α is a weight parameter and is empirically set as 0.6.

## IV. DATASETS

Three OCTA segmentation datasets are used in this work, including two publicly available datasets (ROSE-1 and OCTA_3M) and our published FAROS dataset.

### A. ROSE-1

ROSE-1 is a subset of ROSE [3], and contains 117 OCTA images from 39 subjects including 26 patients with Alzheimer's disease and 13 healthy subjects. The image acquisition equipment is a commercial 70 kHz RTVue-XR Avanti SD-OCT system (Optovue, Inc, Fremont, CA, USA) equipped with AngioVue software. All images were captured in a 3 mm × 3 mm central foveal area with an image resolution of 304 × 304 pixels. We choose the 39 deep-and-superficial (corresponds to from ILM to OPL) angiogram images. Image experts and clinicians graded centerline-level capillaries and pixel-level retinal macrovessel annotations simultaneously and chose their consensus to be the ground truth for RVC, as exemplified in Fig. 3a. For realizing multi-task segmentation, two proficient image experts supplemented pixel-level FAZ labels with labelme software [39].

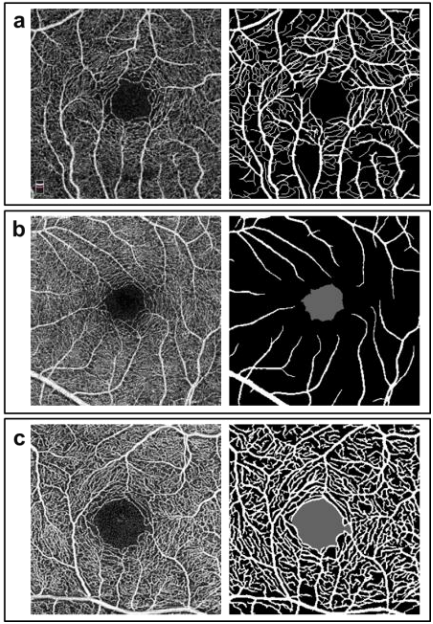

Fig. 3. Examples of three OCTA segmentation datasets used in this paper. **a** ROSE-1 dataset. **b** OCTA_3M dataset. **c** Our released FAROS dataset. In each frame, the left and right images are the original OCTA image and its corresponding ground truth in the dataset, respectively.

### B. OCTA_3M

OCTA_3M is a subset of OCTA-500 [24], and consists of 200 OCTA images with a 3 mm × 3 mm field of view. All images were acquired using a commercial 70 kHz RTVue-XR Avanti SD-OCT system (Optovue, Inc, Fremont, CA, USA) equipped with AngioVue software. To present more detailed retinal information, we choose maximum projected images between limiting membrane (ILM) layer and outer plexiform layer (OPL) with an image resolution of 304 × 304 pixels in our study. RVN with only macrovessels (RVM) and FAZ labels were graded by image experts and clinicians respectively, and the consensus of them was chosen to be the ground truth for multi-task segmentation. An example of an original image and corresponding ground truth in OCTA_3M is shown in Fig. 3b.

### C. FAROS

FAROS dataset consists of 40 OCTA images from 40subjects with 20 eyes from healthy subjects, 5 eyes from patients with DR, 5 eyes from patients with AMD, and 10 eyes from patients with RVO. All images were collected by a commercial 200 kHz ZEISS PlexElite 9000 swept-source OCT angiography (Carl Zeiss Meditec, Inc, Dublin, CA, USA). In accordance with the Declaration of Helsinki, all data were collected with the approval of the relevant authorities and the informed consent of the patients. Each image was resized to a resolution of 304 × 304 pixels, which covered a 3 mm × 3 mm central foveal area.

Since manually annotating capillaries with pixel-level is complex, difficult to manipulate, and quite subjective among different annotators, we employ a semi-automatic scheme to obtain pixel-level masks of our collected OCTA images. As illustrated in Fig. 4, we first trained a multi-task segmentation model based on the ROSE-1 (with additional FAZ labels) dataset and our proposed OMSN, the detail of which is elaborated in the next section. When this model functioned well in the segmentation tasks, we input all the raw OCTA images from FAROS into the trained multi-task segmentation model, obtaining decent RVC and FAZ segmentation masks according to adaptive thresholds. Then five image experts and ocular clinicians were involved in annotating these segmentation masks manually with Adobe Photoshop 2020 software (Adobe System, San Jose, CA, USA), the consensus of whom led to the final fully annotated ground truths (Fig.3c).

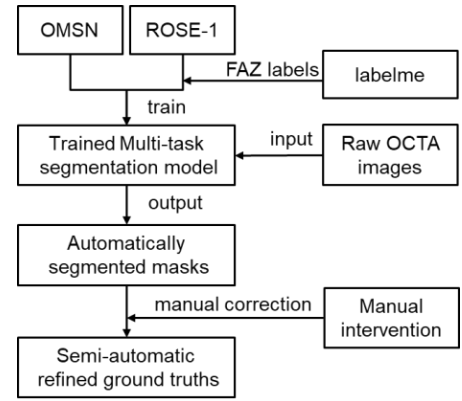

Fig. 4. The flow diagram of the semi-automatic scheme.

## V. EXPERIMENTS AND RESULTS

### A. Experimental Setup

Our proposed OMSN is implemented within the Pytorch framework with a single NVIDIA GPU (GeForce GTX Titan Xp). We choose the adaptive moment estimation as our general learning optimizer with a weight decay of 0.0001. The learning rate is set to 0.0001 initially and satisfies:

$$\text{lr} = \text{lr}_{\text{init}} \times \left(1 - \frac{iter}{\max\_iter}\right)^{power} \tag{10}$$

where $\text{lr}_{\text{init}}$ is the initial learning rate, $iter$ indexes the number of iterations, $\max\_iter$ means the total training epochs, and $power$ is set to 0.9.

Due to limited amount of data, we apply data augmentation by random rotation and contrast enhancement to improve the segmentation performance. All the used datasets are randomly divided into training set, validation set, and testing set in proportion to 6:2:2 respectively. Afterward, we set maximum training epochs to 200, batch size to 2, and start training the model. The early stopping criterion is applied during the training process. Specifically, we calculate the Dice coefficient of the validation set and preserve the best model when the Dice

TABLE I
THE CORRESPONDENCE BETWEEN DATASETS AND SEGMENTATION MODELS

| Task | Segmentations | Datasets | Annotations | Number |
|---|---|---|---|---|
| Single-task | FAZ | OCTA_3M | FAZ | 200 |
| | RVC | ROSE-1 | RVC | 39 |
| Multi-task | RVM and FAZ | OCTA_3M | RVM and FAZ | 200 |
| | RVC and FAZ | ROSE-1 | RVC and FAZ | 39 |
| | RVC and FAZ | FAROS | RVC and FAZ | 40 |

coefficient reaches its maximum.

### B. Experiments Implementation

In this study, we train our proposed OMSN on three OCTA datasets and obtain five different segmentation models, whose correspondences are shown in Table I. Figs. 1b and 1c show the implementation of single-task and multi-task segmentation, respectively. For single-task segmentation, a dataset with single-class annotation (RVC and FAZ) is first input to the OMSN with an output channel equal to 1. Then the predicted images go through a Gaussian adaptive threshold module, obtaining a binary image with local detail optimization. Then, a morphological operation module is applied to adjust the connectivity of RVC or FAZ based on prior criteria. At last, the loss function is calculated and propagated back to supervise the whole process. Differently, the multi-task segmentation dataset has 3-dimensional multiclass annotations (RVM/RVC and FAZ), and is encoded by one-hot to create new binary matrices to match each class in the original data. Correspondingly, OMSN has 3 output channels and performs Gaussian adaptive threshold and morphological operations on each channel.

All the hyper-parameters during the training processes are manually adjusted to achieve the best performance of the segmentation models. For each model, we conduct a comprehensive evaluation and adequate comparative experiments to compare our method with some well-known approaches in terms of both quantity and quality.

### C. Evaluation Metrics

To achieve objective and scientific evaluation of the segmentation performance of the above-mentioned method, we adopt the following metrics for quantitative comparison:

Balanced Accuracy (BACC) = (Sensitivity + Specificity) / 2;

G-mean score (G-means) = $\sqrt{Sensitivity \times Specificity}$;

Dice coefficient (Dice) = 2×TP/ (FP+FN+2×TP);

TABLE II
QUANTITATIVE COMPARISON OF FAZ SEGMENTATION RESULTS ON OCTA_3M DATASET

| Methods | Dice (%) | BACC (%) | JAC (%) | G-means (%) |
|---|---|---|---|---|
| Unet [32] | 96.81±2.96 | 98.69±2.09 | 93.97±5.14 | — |
| Unet++ [40] | 97.10±2.10 | 98.85±1.16 | 94.44±3.89 | — |
| Unet 3+ [41] | 97.45±1.72 | 99.11±1.12 | 95.09±3.20 | — |
| IPN [10] | 95.05±4.79 | 97.68±3.21 | 90.91±7.98 | — |
| IPN V2 [24] | 97.30±2.33 | 98.83±1.64 | 94.84±4.15 | — |
| IPN V2+ [24] | 97.55±2.38 | 98.74±1.74 | 95.32±4.19 | — |
| **OMSN** | **97.94±1.23** | **99.11±0.73** | **96.00±2.31** | **99.10±0.74** |

TABLE III
QUANTITATIVE COMPARISON OF RVC SEGMENTATION ON ROSE-1 DATASET

| Methods | Dice (%) | BACC (%) | JAC (%) | G-means (%) |
|---|---|---|---|---|
| Unet [32] | 70.12 | — | — | 80.50 |
| ResU-Net [42] | 73.09 | — | — | 81.88 |
| CE-Net [43] | 73.00 | — | — | 82.03 |
| DUNet [44] | 74.03 | — | — | 82.13 |
| CS-Net [22] | 74.88 | — | — | 82.63 |
| OCTA-Net [3] | 75.76 | — | — | 82.75 |
| ACRROSS (CS-Net) [35] | 76.8 | - | - | 83.0 |
| ACRROSS (Unet) [35] | **78.8** | - | - | 85.2 |
| VAFF (Single-input) [36] | 71.09 | 81.16 | — | — |
| VAFF (Multi-input) [36] | 76.58 | 84.62- | — | — |
| **OMSN** | 77.80 | **94.70** | **63.94** | **85.38** |

Jaccard index (JAC) = TP / (TP+FP+FN).

where TP refers to true positive, FP refers to false positive, TN represents true negative, FN represents false negative, Sensitivity = TP / (TP+FN), Specificity = TN / (TN+FP).

### D. Single-task Segmentation Results

The construction workflow of single-task segmentation is illustrated in Fig. 1b. First, a dataset with single-class annotations (RVC, FAZ) is input to the OMSN with an output channel equal to 1. Then the predicted images go through a Gaussian adaptive threshold module, obtaining a binary image with local detail optimization. After that, a morphological operation module is applied to adjust the connectivity of RV/RVC or FAZ based on prior criteria. At last, the single-task loss function $\mathcal{L}_{single}$ is calculated and propagated back to supervise the whole process.

*1) Quantitative comparisons:* For FAZ segmentation model on OCTA_3M, we compare with former outstanding segmentation methods, including Unet [32], Unet++ [40], Unet 3+ [41], image projection network (IPN) [10], IPN-V2 [24], and IPN-V2+ [24]. Table II shows that our FAZ segmentation model performs best in BACC, Dice, and JAC metrics. Although the mean BACC of our FAZ segmentation model is equal to the Unet 3+, the smallest standard deviations are achieved with respect to BACC, Dice, and JAC, which confirms the stability and superiority of the proposed method. For the RVC segmentation model, we draw a comparison with previous methods on ROSE-1, covering Unet [32], ResU-net [42], CE-Net [43], DUNet [44], CS-Net [22], OCTA-Net [3], ACRROSS (with CS-Net or Unet) [35], and VAFF (Single-input and Multi-input) [36]. According to Table III, our proposed OMSN outperforms the previous best methods on BACC and G-means and is second only to ACRROSS (Unet) on Dice.

*2) Qualitative comparisons:* We enumerate two typical segmented images to illustrate each model respectively (Fig. 5). From frame I, it is consistent with clinical experience that the FAZ boundaries in patients' OCTA images are less explicit and have more complicated shapes. Therefore, we may obtain poor results in patient cases when we compare the segmented image with the manually annotated ground truth. Nevertheless, as indicated by the green arrow in Fig .5 A1 and A3, the pointed blood vessel appears discontinuous while the manually annotated ground truth recognizes it as continuous, and our model discovers its discontinuity. This vascular discontinuity may result from disruption of vascular structures or projection artefacts, in either case, we need a sufficiently precise method to segment the 2D OCTA projection maps.

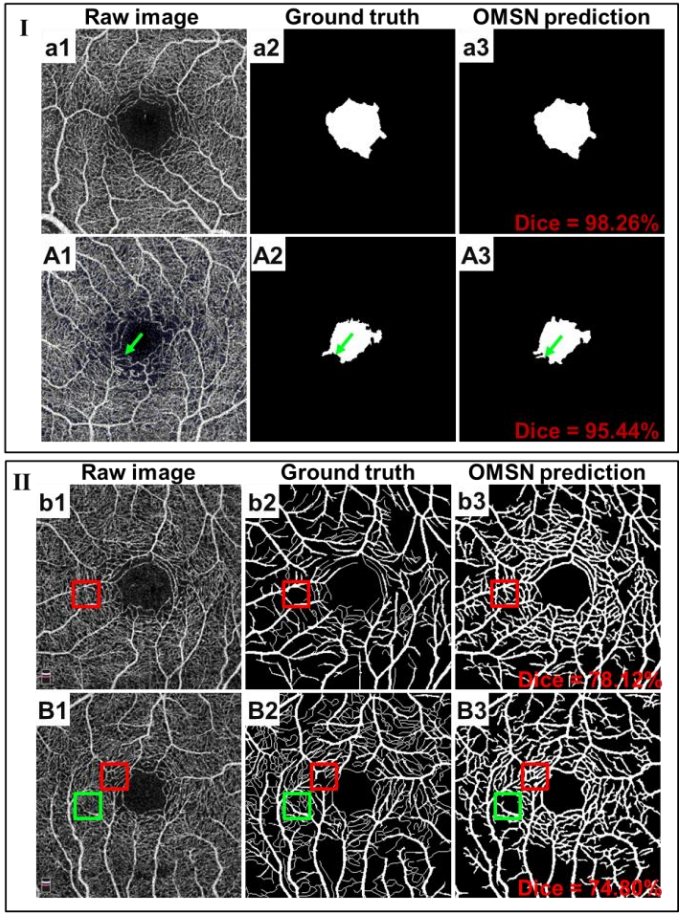

Fig.5. Single-task segmentation results on OCTA_3M and ROSE-1. **I** a1-a3 and A1-A3 illustrate two FAZ segmentation instances on OCTA_3M respectively. **II** b1-b3 and B1-B3 show two instances of RVC segmentation on ROSE-1. B1 shows obvious motion artifacts at the bottom.

From frame II, we find that the bottom of B1 appears obvious motion artefacts that may come from the image acquisition process. It is simple to discover that the better image quality is, the better segmentation performs. Furthermore, we find that the red-boxed areas are not annotated in the ground truth, but segmented using our model. In comparison, the green-boxed area is labeled in the ground truth, while our model can hardly recognize it. Any adjustment in the post-processing process is not beneficial to this problem. Moreover, the motion artefacts in the ROSE-1 dataset make it unrepresentative for universal OCTA images analysis.

### E. Multi-task Segmentation Results

The construction workflow of multi-task segmentation is illustrated in Fig. 1c. First, a dataset with 3-dimensional multi-class annotations (RVM/RVC and FAZ) is encoded by one-hot encoding, creating new binary matrices to match each class in the original data. The dataset is then passed through the OMSN with an output channel equal to 3. After that, channel-wise Gaussian adaptive threshold and morphological operations are implemented to optimize the results for every class respectively. Finally, the multi-task loss function $\mathcal{L}_{multi}$ is computed and propagated back to supervise the entire process.

*1) Quantitative comparisons:* We implement concise ablation studies by comparing the OMSN with Unet, Unet 3+, and ResNeSt-backboned Unet, and the results of multi-task segmentation models are shown in Table IV. For RVM and FAZ segmentation on OCTA_3M, our proposed OMSN outperforms the other compared methods in Dice, BACC, JAC and G-means metrics. More importantly, when we compare the results of the single-task segmentation models on OCTA_3M in Table II, the Dice and JAC of the multi-task segmentation are significantly better than that of the single-task segmentation. Although BACC and G-means are slightly lower than that of single-task segmentation, but have smaller standard deviations. For RVC and FAZ segmentation on ROSE-1, our proposed OMSN achieves better all metrics than the compared methods. Comparing with the single-task RVC segmentation model (Table III), we find that Dice and JAC metrics are lifted significantly while the BACC is reduced slightly. Numerically speaking, we can still draw a similar conclusion that our proposed OMSN performs better in multi segmentation tasks than in single segmentation tasks. For RVC and FAZ segmentation on FAROS, our proposed OMSN also surpasses the other three methods. When we longitudinally compare results for the two RVC and FAZ segmentation models, we find that whatever method is used, all metrics appear larger in FAROS than in ROSE. Therefore, we can preliminarily conclude that our proposed FAROS dataset is more reliable than the ROSE-1 dataset in segmentation tasks.

*2) Qualitative comparisons:* To qualitatively evaluate the multi-task performance of the compared methods, we enumerate an example from OCTA_3M, ROSE-1, and FAROS,

TABLE IV
QUANTITATIVE COMPARISON OF MULTI-TASK SEGMENTATION RESULTS ON THREE DATASETS

| Methods | RVM and FAZ segmentation on OCTA_3M | | | | RVC and FAZ segmentation on ROSE-1 | | | | RVC and FAZ segmentation on FAROS | | | |
|---|---|---|---|---|---|---|---|---|---|---|---|---|
| | Dice (%) | BACC (%) | JAC (%) | G-means (%) | Dice (%) | BACC (%) | JAC (%) | G-means (%) | Dice (%) | BACC (%) | JAC (%) | G-means (%) |
| Unet [32] | 96.51± 0.68 | 97.46± 0.39 | 93.28± 1.27 | 97.46± 0.39 | 87.90± 4.95 | 89.62± 3.80 | 78.74± 7.50 | 89.61± 3.83 | 91.33± 0.96 | 93.82± 0.67 | 84.07± 1.64 | 93.82± 0.67 |
| Unet 3+ [41] | 98.57± 0.27 | 98.66± 0.20 | 97.09± 0.52 | 98.99± 0.19 | 90.17± 2.43 | 92.67± 1.83 | 82.33± 4.00 | 92.90± 1.84 | 91.79± 0.89 | 94.08± 0.64 | 84.99± 1.54 | 94.28± 0.64 |
| ResNeSt-Unet | 98.19± 0.31 | 98.97± 0.24 | 96.46± 0.61 | 98.97± 0.24 | 89.96± 2.40 | 92.53± 1.82 | 81.83± 3.92 | 92.53± 1.82 | 92.08± 0.99 | 94.27± 0.72 | 85.48± 1.73 | 94.46± 0.72 |
| OMSN | **98.76± 0.22** | **99.02± 0.17** | **97.55± 0.44** | **99.02± 0.17** | **90.48± 2.50** | **93.54± 1.88** | **82.73± 4.11** | **93.52± 1.89** | **92.52± 0.96** | **94.64± 0.69** | **86.11± 1.68** | **94.63± 0.69** |

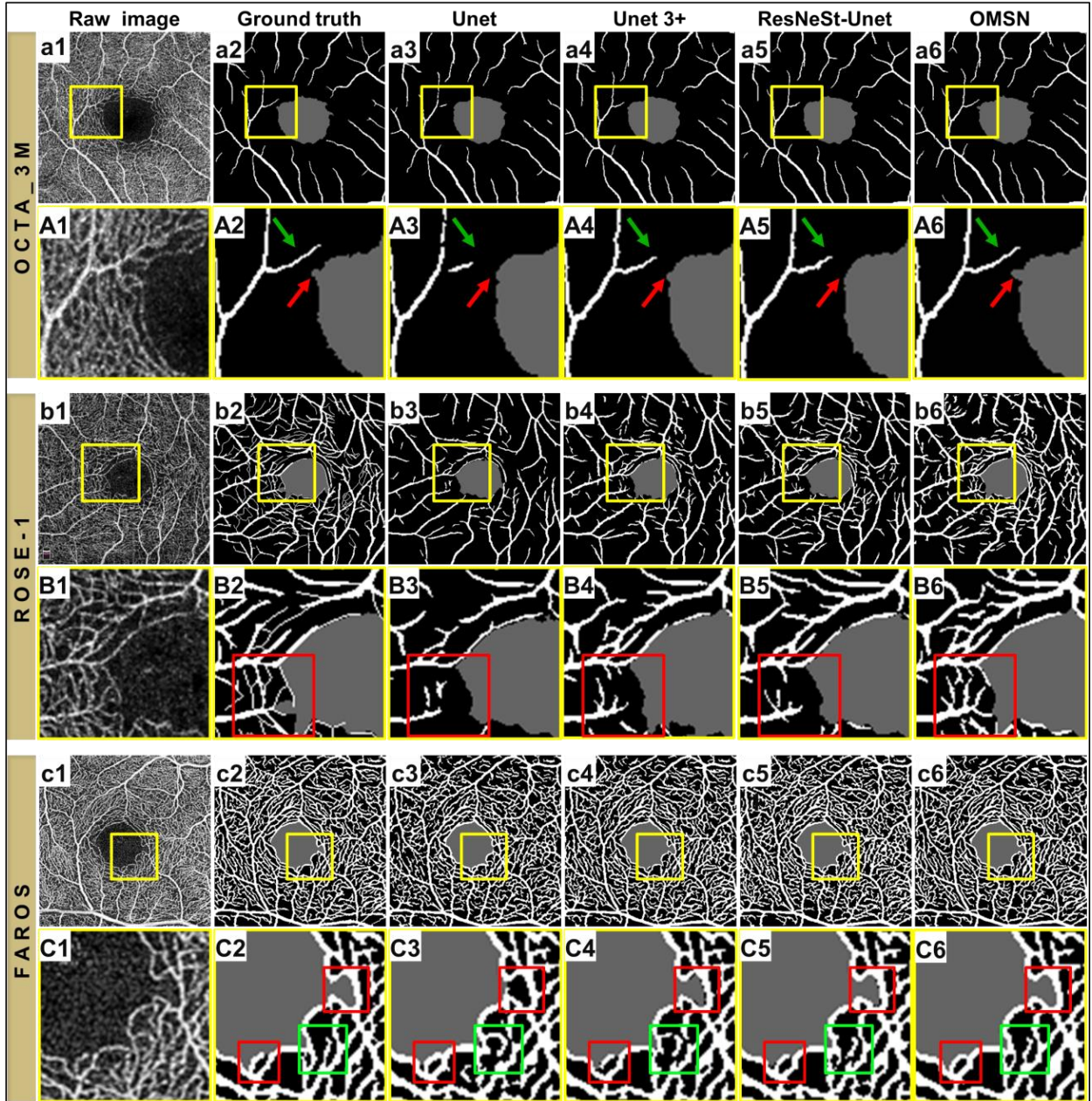

Fig.6. Multi-task segmentation results on three datasets. a1-a2, b1-b2, and c1-c2 are representative paired examples from OCTA_3M, ROSE-1, and FAROS respectively. Note that A1-A6, B1-B6, and C1-C6 are the corresponding enlarged areas of a1-a6, b1-b6 and c1-c6

respectively, for intuitive explanation (Fig. 6). In all examples, Unet performs relatively poorly due to its poor capability of recognizing tiny capillaries and intricately shaped FAZ edges. Unet 3+ and ResNeSt-Unet have both identified respectable small capillaries, while Unet 3+ performs better than ResNeSt-Unet in distinguishing FAZ borders. Nevertheless, none of these methods can realize fine capillaries and FAZ borders segmentation. In contrast, our proposed OMSN can recognize more visually valuable information. Specifically, in Fig.6 A2-A6, the red arrows indicate that RV recognized by OMSN are more legible, successive, and explicit in all the minor details, and the green arrows manifest that the boundary of the FAZ segmented by OMSN matches more with the ground truth in locations with tiny bumps and slight deformation. Similarly, in Fig.6 B2-B6, our proposed OMSN can recognize conspicuously more retinal vascular networks and more explicit FAZ boundaries in the red-boxed areas. In Fig.6 C2-C6, the red-boxed areas indicate that Unet performs poorly in recognizing FAZ edges with complex shapes, Unet 3+ and ResNeSt-Unet perform similarly and relatively better than Unet in FAZ segmentation, and our proposed OMSN best matches the ground truth in identifying FAZ edges. As for the segmentation of capillaries, the green-boxed areas reflect that Unet, Unet 3+, and ResNeSt-Unet all have the difficulty in distinguishing some tiny capillaries which results in the discontinuity of the segmented capillaries, while our proposed OMSN pays more attention to the areas of capillaries with precise labels, thus obtaining more consecutive segmentation results of capillaries. Furthermore, comparing the instances from ROSE-1 and FAROS, clearly many more capillaries are recognized, which further proves the superiority of the FAROS dataset.

## VI. CONCLUSIONS

In this paper, we introduce an innovative deep network OMSN to achieve accurate OCTA microstructure segmentation. The network architecture mainly consists of a multi-scale skip connection encoder-decoder framework and split-attention-based ResNeSt blocks. The multi-scale skip connections make OMSN focus more attention on visual features of the middle and deep network layers. The introduction of ResNeSt blocks not only boosts the network performance by improving the learned feature representations, but also takes into account the computation efficiency. Based on OMSN and two publicly available OCTA datasets, we develop two independent single-task segmentation models (RVN and FAZ). Thorough evaluation and comparative experiments demonstrate that the OMSN achieves ouststanding RVN and FAZ segmentation performance on all used OCTA datasets. Besides, we groundbreakingly build and release FAROS, a fully-labeled OCTA dataset, containing pixel-level RVC and FAZ annotations annotated by a semi-automatic method, which fills the vacancy of the OCTA datasets field. For the first time, we achieve multi-task OCTA segmentation (RVM/RVC and FAZ) based on the two publicly available datasets and FAROS dataset. Moreover, we quantitatively demonstrate that the concurrent RVM/RVC and FAZ segmentation models outperform any single-task segmentation model built on the same dataset.

Although our method achieves excellent segmentation performance, there are some limitations. First, like the previous method, our proposed OMSN still has trouble recognizing tiny capillary areas with low image quality. In future research, we will dabble in this field of recognizing low-quality OCTA images with image registration [45]. Then, although our published FAROS dataset has better image quality than the former publicly available OCTA dataset, it still possesses the defect of an insufficient number of images like most medical image segmentation datasets. Besides, the raw OCTA images captured by ZEISS PLEX Elite 9000 have an extremely high resolution of 1024×1024, which means much more elaborate information. In the future, we will first consider including more data from patients and healthy subjects to enrich the FAROS dataset and establish a universal diagnostic model by analyzing the differences in RVN and FAZ related parameters of different diseases. Furthermore, we intend to put more human and material resources into building a fully annotated 1024×1024 high-resolution OCTA dataset. In this case, all the outstanding automatic segmentation methods will have the opportunity to be applied to this advanced imaging technique, thus providing a powerful boost to artificial intelligence diagnosis.